\documentclass[12pt]{article}
\usepackage{amssymb}
\usepackage{amsmath}
\usepackage{theorem}

\newtheorem{theorem}{Theorem}[section]

\newtheorem{proposition}[theorem]{Proposition}
\newtheorem{corollary}[theorem]{Corollary}
\newtheorem{remark}[theorem]{Remark}
\begin{document}
\author{Luigi Accardi$\dagger $ and Masanori Ohya$\ddagger $ \\
$\dagger$ Graduate School of Polymathematics,\\
Nagoya University, Chi\-ku\-sa--ku,\\
Na\-go\-ya, 464--01, Japan,\\
and Centro V. Volterra, Universit\`{a} degli Studi di\\
Roma ``Tor Vergata'' -- 00133 Rome, Italy\\
E-mail: accardi@volterra.mat.uniroma2.it,\\
and accardi@math.nagoya-u.ac.jp\\
$\ddagger $Department of Information Sciences\\
Science University of Tokyo\\
Noda City, Chiba 278-8510, Japan\\
E-mail: ohya@is.noda.sut.ac.jp}
\title{TELEPORTATION OF\ GENERAL\ QUANTUM\ STATES }
\date{}
\maketitle

\footnotetext{
\noindent {Invited talk to the: International Conference on quantum
information and computer, Meijo University 1998}}

\section{Introduction}

Quantum teleportation has been introduced by Benett et al. \cite{BBCJPW}
and discussed by a number of authors in the framework of the singlet state \cite{BBPSSW}. Recently, a rigorous formulation of the teleportation problem of arbitrary quantum states by means of quantum channel was given in \cite{IOS} based on the general channel theoretical formulation of the quantum gates introduced in \cite{OW}. In this note we discuss a generalization of the scheme proposed in \cite{IOS} and we give a general method to solve the teleportation problem in spaces of arbitrary finite dimensions.

\section{Formulation of the problem}

The set of all quantum states on a Hilbert space ${\cal H}$, identified
to the set of the density operators, is denoted by ${\cal S}({\cal H}), $ namely, 
\[
{\cal S}({\cal H})\equiv \{\rho \in B({\cal H})\ ;\ \rho ^{*}=\rho
,\ \rho \geq 0,\ \mathrm{tr}\rho =1\} 
\]
where $B({\cal H})$ is the set of all bounded operators on ${\cal H}$.

The following is a generalization of the channel theoretical approach to the
teleportation problem proposed by \cite{IOS}:

\begin{itemize}
\item[{STEP 0}]  : Alice has a unknown quantum state $\rho ^{(1)}$ on a
Hilbert space ${\cal H}_{1}$, and she wants to teleport it to Bob.

\item[{STEP 1}]  :Two auxiliary Hilbert spaces ${\cal H}_{2}$ and ${\cal H}_{3}$, attached to Alice and to Bob respectively, are introduced.

One fixes a set of (entangled) states

\begin{equation}
\sigma _{k}^{(23)}\in {\cal S}\left( {\cal H}_{2}\otimes {\cal H}_{3}\right)   \label{2.1}
\end{equation}
in the (Alice, Bob)--space, having certain prescribed correlations and one
prepares an ensemble of the combined system $\left( 1,2,3\right) $ in the
state
\end{itemize}

\begin{equation}
\rho _{k}^{\left( 123\right) }\equiv \rho ^{\left( 1\right) }\otimes \sigma
_{k}^{23}  \label{2.2}
\end{equation}
on the space ${\cal H}_{1}\otimes {\cal H}_{2}\otimes {\cal H}_{3}$

\begin{itemize}
\item[{STEP 2}]  : One then fixes a family of mutually orthogonal
projections 
\begin{equation}
\left\{ F_{k}^{(12)}\right\}   \label{2.3}
\end{equation}
on the Hilbert space ${\cal H}_{1}\otimes {\cal H}_{2}$, corresponding
to an observable $A:=\sum_{k}\lambda _{k}F_{k}^{(12)}$ and having fixed one index $k$, Alice performs a first kind incomplete measurement, involving only the $(1,2)$ system, which filters the (arbitrarily chosen) value $\lambda_{k}$, i.e. after the measurement on the given ensemble (\ref{2.2}), of identically prepared systems, only those with 
$A=\lambda_{k}$ are allowed to pass. According to quantum mechanics, after Alice's measurement, the state of the $(1,2,3)$ system becomes 
\begin{equation}
\rho _{k}^{(123)}:={\frac{\left( F_{k}^{(12)}\otimes 1_{3}\right) \rho
_{k}^{(123)}\left( F_{k}^{(12)}\otimes 1_{3}\right) }{\mathrm{tr}_{123}\left( F_{k}^{(12)}\otimes 1_{3}\right) \rho _{k}^{(123)}\left(
F_{k}^{(12)}\otimes 1_{3}\right) }}  \label{2.4}
\end{equation}
where $\mathrm{tr}_{123}$ is the full trace on the Hilbert space ${\cal H}_{1}\otimes {\cal H}_{2}\otimes {\cal H}_{3}.$

\item[{STEP 3}]  : Bob is informed which measurement was done by Alice. This
is equivalent to transmit the information that the $k$--th eigenvalue was
chosen. This information is transmitted from Alice to Bob without
disturbance and by means of completely classical tools (e.g. telephone).

\item[{STEP 4}]  : By making partial measurements on the system $3$, that
is, on the system corresponding to the auxiliary space related to him and
not to the original system, Bob can obtain the state $\rho ^{(3)}$ induced
by the state (\ref{2.4}) and reconstruct the state $\rho ^{(1)}$ on the
system $1$by unitary keys provided to him. Notice that this state $\rho
^{(3)}$is unknown by Alice unless the ensemble (\ref{2.2}) has been prepared
by her.
\end{itemize}

\medskip

Here we must distinguish two cases:

\begin{itemize}
\item[{(I)}]  Bob can perform his experiments on the same ensemble of
systems found by Alice as a result of her measurement

\item[{(II)}]  Bob and Alice are spatially separated so that the situation
of case (I) is not realizable. In this case Bob has to prepare an ensemble
of systems in the state $\rho ^{(123)}$ and therefore also this state has to
be transmitted by Alice by means of classical communication.
\end{itemize}

The crucial point of the construction is that, knowing the information
transmitted by Alice about which measurement was done by her, Bob is able to
reconstruct in a unique way the original state $\rho ^{(1)}$, of system $1$,
from the state $\rho ^{(3)}$, of system $3$. \medskip When the state $\sigma
_{k}^{(23)}$ is independent of $k$, the above problem reduces to the channel
theoretical formulation of \cite{IOS}. \medskip The above procedure can be
realized by a channel (dual of a completely positive map) 
\[
\Lambda _{k}^{*}:{\cal S}\left( {\cal H}_{1}\right) \longrightarrow 
{\cal S}\left({\cal H}_{3}\right) 
\]
composed of the following four channels:

\begin{description}
\item[(i)]  A trivial (i.e. product) lifting in the sense of \cite{AO} 
\begin{equation}
\gamma _{k}^{*}:{\cal S}\left( {\cal H}_{1}\right) \longrightarrow 
{\cal S}\left( {\cal H}_{1}\otimes {\cal H}_{2}\otimes {\cal H}_{3}\right)   \label{2.5}
\end{equation}

\begin{equation}
\gamma _{k}^{*}\left( \rho ^{(1)}\right) =\rho ^{(1)}\otimes \sigma
_{k}^{(23)},\qquad \forall \rho ^{(1)}\in {\cal S}\left( {\cal H}_{1}\right)   \label{2.6}
\end{equation}
expresses the independent coupling of the initial state $\rho ^{(1)}$ with
the state $\sigma _{k}^{(23)}$.

\item[(ii)]  The second step is described by a measurement type channel 
\begin{equation}
\pi _{k}^{*}:{\cal S}\left( {\cal H}_{1}\otimes {\cal H}_{2}\otimes 
{\cal H}_{3}\right) \longrightarrow {\cal S}\left( {\cal H}_{1}\otimes {\cal H}_{2}\otimes {\cal H}_{3}\right)   \label{2.7}
\end{equation}
of the form 
\[
\pi _{k}^{*}\left( \rho _{k}^{(123)}\right) :={\frac{\left(
F_{k}^{(12)}\otimes 1_{3}\right) \rho _{k}^{(123)}\left( F_{k}^{(12)}\otimes
1_{3}\right) }{\mathrm{tr}_{123}\left( F_{k}^{(12)}\otimes 1_{3}\right) \rho
_{k}^{(123)}\left( F_{k}^{(12)}\otimes 1_{3}\right) },}
\]
where $\rho _{k}^{(123)}\in {\cal S}\left( {\cal H}_{1}\otimes 
{\cal H}_{2}\otimes {\cal H}_{3}\right) $, corresponding to an
incomplete first kind measurement describing the state change determined by
Alice's filtering of the eigenvalue $\lambda _{k}$ of the observable $A$

\item[(iii)]  The third step is defined by the channel $a^{*}:{\cal S}\left( {\cal H}_{1}\otimes {\cal H}_{2}\otimes {\cal H}_{3}\right)
\longrightarrow {\cal S}\left( {\cal H}_{3}\right) $ defined by 
\[
\rho _{k}^{(3)}=a^{*}\left( \rho _{k}^{(123)}\right) =\mathrm{tr}_{12}\rho
_{k}^{(123)},\qquad \forall \rho _{k}^{(123)}\in {\cal S}\left( {\cal H}_{1}\otimes {\cal H}_{2}\otimes {\cal H}_{3}\right) .
\]
Here $\mathrm{tr}_{12}$ is the partial trace on the Hilbert space ${\cal H}_{1}\otimes {\cal H}_{2}$ 
\[
<\Phi _{1},\mathrm{tr}_{12}\,Q\Phi _{2}>\,\,\equiv \sum\limits_{n}<\Psi
_{n}\otimes \Phi _{1},\,Q\Psi _{n}\otimes \Phi _{2}>,\;\; Q\in B({\cal H}_{1}\otimes {\cal H}_{2}\otimes {\cal H}_{3})
\]
for any CONS $\{\Psi _{n}\}$ of ${\cal H}_{1}\otimes {\cal H}_{2}$ and
any $\Phi _{1},\Phi _{2}\in {\cal H}_{3}$. This channel $a^{*}$
corresponds to Bob's partial measurement over the system $3$ and describes
the reduction, from the state $\rho ^{(123)}$ obtained after Alice's
measurement, to the state $\rho ^{(3)}$, obtained by Bob. Thus the whole
teleportation process above is written by the channel 
\[
\Lambda _{k}^{*}:{\cal S}\left( {\cal H}_{1}\right) \longrightarrow 
{\cal S}\left( {\cal H}_{3}\right) 
\]
\[
\Lambda _{k}^{*}\equiv \Lambda _{k,A\to B}^{*}\equiv a^{*}\circ \pi_{k}^{*}\circ \gamma ^{*}
\]

The above subscript ``$k$'' means that the channels $\Lambda _{k}^{*}$ and $\Lambda _{k,A\to B}^{*}$ depend on the choice of Alice's measurement $F_{k}^{(12)}$. More precisely, $\forall \rho ^{(1)}\in {\cal S}\left( 
{\cal H}_{1}\right) $ 
\begin{equation}
\Lambda _{k}^{*}\rho ^{(1)}\equiv \mathrm{tr}_{12}\left[ {\frac{\left(
F_{k}^{(12)}\otimes 1_{3}\right) \left( \rho ^{(1)}\otimes \sigma
_{k}^{(23)}\right) \left( F_{k}^{(12)}\otimes 1_{3}\right) }{\mathrm{tr}_{123}\left( F_{k}^{(12)}\otimes 1_{3}\right) \left( \rho ^{(1)}\otimes
\sigma _{k}^{(23)}\right) \left( F_{k}^{(12)}\otimes 1_{3}\right) }}\right] 
\label{2.8}
\end{equation}
Note that the channel $\Lambda _{k}^{*}$ is generally non linear.
\end{description}

With these notations we can formulate the general mathematical problem of
teleportation as follows. \bigskip

\bigskip Given the initial Hilbert space ${\cal H}_{1}$, find:

\begin{itemize}
\item[{(1)}]  two auxiliary Hilbert spaces ${\cal H}_{2}$, ${\cal H}_{3}$

\item[{(2)}]  a family of entangled states $\sigma _{k}^{(23)}$ on ${\cal H}_{2}\otimes {\cal H}_{3}$

\item[{(3)}]  a family of mutually orthogonal projections $\left\{
F_{k}^{(12)}\right\} $ acting on ${\cal H}_{1}\otimes {\cal H}_{2}$

\item[{(4)}]  for each $k$ a unitary operator $U_{k}$ such that the
associated unitary channel 
\[
u_{k}^{*}:{\cal S}\left( {\cal H}_{3}\right) \longrightarrow {\cal S}\left( {\cal H}_{1}\right) 
\]
\[
u_{k}^{*}\left( \rho ^{(3)}\right) =U_{k}\rho ^{(3)}U_{k}^{*}\quad \forall
\rho ^{(3)}\in {\cal S}\left( {\cal H}_{3}\right) 
\]
so that it satisfies the identity 
\begin{equation}
\Lambda _{k}^{*}\rho ^{(1)}=U_{k}^{*}\rho ^{(1)}U_{k}  \label{2.9}
\end{equation}
for any $k$ and for any state $\rho ^{(1)}\in {\cal S}({\cal H}_{1})$
or at least for $\rho ^{(1)}$ in a preassigned subset of ${\cal S}({\cal H}_{1})$. \bigskip 
\end{itemize}

If the conditions (2), (3), (4)above are replaced by the weaker ones:

\begin{itemize}
\item[{(2')}]  a single entangled state $\sigma ^{(23)}$acting on ${\cal H}_{1}\otimes {\cal H}_{2}$

\item[{(3')}]  a single projection $F^{(12)}$ acting on ${\cal H}_{1}\otimes {\cal H}_{2}$

\item[{(4')}]  a single unitary operator $U$ such that the identity 
\begin{equation}
\Lambda ^{*}\rho ^{(1)}=U^{*}\rho ^{(1)}U  \label{2.10}
\end{equation}
holds for any state $\rho ^{(1)}\in {\cal S}({\cal H}_{1})$ and the
channel determined by $\sigma ^{(23)}$ and $F^{(12)},$ \bigskip then we
speak of the \mbox{weak teleportation problem}.
\end{itemize}

The connection between the weak and the general teleportation problem is the
following. Given a family $\{\sigma _{k}^{(23)},F_{k}^{(12)},U_{k}\}$ of
solutions of the weak teleportation problem for each k such that the
projections $F_{k}^{(12)}$ are mutually orthogonal, then this family
provides a solution of the general teleportation problem. In the following
section, we shall solve the weak teleportation problem, and then we shall
use this result to solve the general teleportation problem.

\section{Solution of the weak teleportation problem}

In the notations of the previous section, we shall assume that 
\[
N=\dim {\cal H}_{1}<+\infty 
\]
Under this assumption we shall look for a solution of the weak teleportation
problem in which 
\[
N=\dim {\cal H}_{1}=\dim {\cal H}_{2}=\dim {\cal H}_{3}
\]
\begin{equation}
\sigma ^{(23)}=|\psi \rangle \langle \psi |  \label{3.1}
\end{equation}
\begin{equation}
F:=|\xi \rangle \langle \xi |  \label{3.2}
\end{equation}
where $\psi \in {\cal H}_{2}\otimes {\cal H}_{3}$ and $\xi \in 
{\cal H}_{1}\otimes {\cal H}_{2}$ are unit vectors. In the following
we identify $F$ with 
\begin{equation}
F=|\xi \rangle \langle \xi |\otimes 1_{3}\in \Pr oj({\cal H}_{1}\otimes 
{\cal H}_{2}\otimes {\cal H}_{3})  \label{3.3}
\end{equation}
and we look for a unitary transformation $U:{\cal H}_{3}\to {\cal H}_{1}$ such that for any density matrix $\rho \in {\cal S}({\cal H}_{1})
$ one has 
\begin{equation}
U\cdot {\frac{tr_{12}(F(\rho \otimes |\psi \rangle \langle \psi |)F)}{tr(F(\rho \otimes |\psi \rangle \langle \psi |)F)}}U^{*}\,=\rho   \label{3.4}
\end{equation}

Under these assumptions, let us fix three arbitrary orthonormal bases: 
\begin{equation}
(\varepsilon _{\alpha })\qquad \mbox{of}\qquad {\cal H}_{3}  \label{3.5}
\end{equation}

\begin{equation}
(\varepsilon _{h}^{\prime })\qquad \mbox{of}\qquad {\cal H}_{2}
\label{3.6}
\end{equation}

\begin{equation}
(\varepsilon _{n}^{\prime \prime })\qquad \mbox{of}\qquad {\cal H}_{1}
\label{3.7}
\end{equation}

\bigskip

\begin{proposition}
In the notations and assumptions of this section, fix
an arbitrary $N\times N$ complex unitary matrix $(\lambda _{\gamma \alpha })$
and define 
\begin{equation}
\psi :=\sum \lambda _{h\alpha }|\varepsilon _{h}^{\prime }\rangle \otimes
|\varepsilon _{\alpha }\rangle \in {\cal H}_{2}\otimes {\cal H}_{3}
\label{3.8}
\end{equation}
\begin{equation}
\xi ={\frac{1}{N^{1/2}}}\,\sum_{\mu }\varepsilon _{\mu }^{\prime \prime
}\otimes \varepsilon _{\mu }^{\prime }  \label{3.9}
\end{equation}
\end{proposition}

Then if $U:{\cal H}_{3}\to {\cal H}_{1}$ is the unique unitary
operator such that 
\begin{equation}
\sum_{h}\overline{\lambda }_{h\alpha }\varepsilon _{h}^{\prime \prime
}=U\varepsilon _{\alpha }  \label{3.10}
\end{equation}
(existing because of our assumption on the $\lambda _{k\beta }$'s), the
triple $(\psi ,\xi ,U)$ satisfies 
\begin{equation}
tr_{12}(F(\rho \otimes |\psi \rangle \langle \psi |)F)={\frac{1}{N}}U^{*}\rho U  \label{3.11}
\end{equation}
for any choice of the density operator $\rho \in {\cal S}({\cal H}_{1})
$ and with $F$ given by (\ref{3.2}), (\ref{3.3}). 

\bigskip

\noindent 
{\bf Proof.} Notice that under the conditions (\ref{3.8}) and (\ref{3.9}), we use in a crucial way the finite dimensionality of ${\cal H}_{2}$ and ${\cal H}_{3}$. In particular 
\begin{equation}
\Vert \psi \Vert ^{2}=\sum_{h,\alpha }|\lambda _{h\alpha
}|^{2}=\sum_{h}1=\dim {\cal H}_{2}=\dim {\cal H}_{3}  \label{3.12}
\end{equation}

We do not normalize $\psi $ because in all the formulae the corresponding
rank one projection will enter both in the numerator and in the denominator.
For $F$ as in (\ref{3.3}), one has, using from now on the convenction of
summation over repeated indices 
\[
F={\frac{1}{N}}|\varepsilon _{\mu }^{\prime \prime }\rangle \langle
\varepsilon _{\mu ^{\prime }}^{\prime \prime }|\otimes |\varepsilon _{\mu
}^{\prime }\rangle \langle \varepsilon _{\mu ^{\prime }}^{\prime }|\otimes
1_{3},
\]
therefore 
\[
F(\rho \otimes |\psi \rangle \langle \psi |)F=\lambda _{h\alpha }\overline{\lambda }_{k\beta }F(\rho \otimes |\varepsilon _{h}^{\prime }\rangle \langle
\varepsilon _{k}^{\prime }|)F\otimes |\varepsilon _{\alpha }\rangle \langle
\varepsilon _{\beta }|=
\]
\[
={\frac{1}{N^{2}}}\,|\varepsilon _{\mu }^{\prime \prime }\rangle \langle
\varepsilon _{\mu ^{\prime }}^{\prime \prime },\rho \varepsilon _{\nu
}^{\prime \prime }\rangle \langle \varepsilon _{\nu ^{\prime }}^{\prime
\prime }|\otimes |\varepsilon _{\mu }^{\prime }\rangle \langle \varepsilon
_{\mu ^{\prime }}^{\prime },\varepsilon _{h}^{\prime }\rangle \langle
\varepsilon _{k}^{\prime },\varepsilon _{\nu }^{\prime }\rangle \langle
\varepsilon _{\nu ^{\prime }}^{\prime }|\otimes |\varepsilon _{\alpha
}\rangle \langle \varepsilon _{\beta }|=
\]
\[
={\frac{1}{N^{2}}}\,\lambda _{h\alpha }\overline{\lambda }_{k\beta }\delta
_{\mu ^{\prime },h}\delta _{k,\nu }\langle \varepsilon _{\mu ^{\prime
}}^{\prime \prime },\rho \varepsilon _{\nu }^{\prime \prime }\rangle
|\varepsilon _{\mu }^{\prime \prime }\rangle \langle \varepsilon _{\nu
^{\prime }}^{\prime \prime }|\otimes |\varepsilon _{\mu }^{\prime }\rangle
\langle \varepsilon _{\nu ^{\prime }}^{\prime }|\otimes |\varepsilon
_{\alpha }\rangle \langle \varepsilon _{\beta }|=
\]
\[
={\frac{1}{N^{2}}}\,\lambda _{h\alpha }\overline{\lambda }_{k\beta }\langle
\varepsilon _{h}^{\prime \prime },\rho \varepsilon _{k}^{\prime \prime
}\rangle |\varepsilon _{\mu }^{\prime \prime }\rangle \langle \varepsilon
_{\nu ^{\prime }}^{\prime \prime }|\otimes |\varepsilon _{\mu }^{\prime
}\rangle \langle \varepsilon _{\nu ^{\prime }}^{\prime }|\otimes
|\varepsilon _{\alpha }\rangle \langle \varepsilon _{\beta }|
\]

Taking the trace over ${\cal H}_{1}\otimes {\cal H}_{2},$ this gives 
\[
={\frac{1}{N}}\lambda _{h\alpha }\overline{\lambda }_{k\beta }\langle
\varepsilon _{h}^{\prime \prime },\rho \varepsilon _{k}^{\prime \prime
}\rangle |\varepsilon _{\alpha }\rangle \langle \varepsilon _{\beta }|={\frac{1}{N}}\langle \overline{\lambda }_{h\alpha }\varepsilon _{h}^{\prime
\prime },\rho \overline{\lambda }_{k\beta }\varepsilon _{k}^{\prime \prime
}\rangle |\varepsilon _{\alpha }\rangle \langle \varepsilon _{\beta }|
\]
Thus, with the choice of $U$ given by (10), (2) becomes 
\[
tr_{12}(F(\rho \otimes |\psi \rangle \langle \psi |)F)={\frac{1}{N}}\langle
\varepsilon _{\alpha },U^{*}\rho U\varepsilon _{\beta }\rangle |\varepsilon_{\alpha }\rangle \langle \varepsilon _{\beta }|={\frac{1}{N}}U^{*}\rho U
\]
which is (\ref{3.11}). 

\section{Uniqueness of the key}

In this section we discuss the uniqueness of the key uintary operator.

\begin{proposition}
Let $\rho=\sum p_{\gamma}P_{\gamma}$ be the spectral decomposition of $\rho \in {\cal H}_{1}$. Then if $U$ and $V$ satisfy equation
 (\ref{2.9}) with the above $\rho$, then there exists a unitary operator $W$ from ${\cal H}_{1}$ to ${\cal H}_{1}$ such that $VU^{*}=\Sigma W_{\gamma }$ with $W_{\gamma }\equiv P_{\gamma }WP_{\gamma }$. Moreover, the equality $W_{\gamma }W_{\gamma }^{*}=\delta _{\gamma \gamma ^{\prime }}P_{\gamma}$ is satisfied.
\end{proposition}

{\bf Proof.} Suppose $U$ and $V$ are two solutions of equation $1_{3}$,
then 
\[
U^{*}\rho U=V^{*}\rho V
\]
or, equivalently 
\begin{equation}
VU^{*}\rho =\rho VU^{*}  \label{4.1}
\end{equation}
This means that $VU^{*}\equiv W:{\cal H}_{1}\to {\cal H}_{1}$ is in
the commutant of $\rho $. Since $W$ is a unitary operator commuting with $\rho$, 
\[
WP_{\gamma }=P_{\gamma }W
\]
is satisfied. Therefore 
\[
W=\sum P_{\gamma }WP_{\gamma }=\sum W_{\gamma }:W_{\gamma }P_{\gamma
}=P_{\gamma }W_{\gamma }=W_{\gamma }.
\]
The equalities 
\[
1=W^{*}W=\sum_{\gamma \gamma ^{\prime }}W_{\gamma }^{*}W_{\gamma ^{\prime
}}=\sum_{\gamma \gamma ^{\prime }}W_{\gamma }^{*}P_{\gamma }P_{\gamma
^{\prime }}W_{\gamma }=\sum_{\gamma }W_{\gamma }^{*}W_{\gamma }.
\]
imply 
\[
W_{\gamma }W_{\gamma ^{\prime }}^{*}=\delta _{\gamma \gamma ^{\prime
}}P_{\gamma }.\;\;\hfill 
\]
\bigskip 

\begin{corollary}
\noindent Let ${\cal H}$ be an Hilbert space of arbitrary dimensions. If $U$ and $V$ are two solutions of the weak teleportation problem corresponding
to the same $F$ and $\psi ,$ then they coincide up to multiplication by a
number of modulus one. 
\end{corollary}

\bigskip

\noindent 
{\bf Proof.} Equation (\ref{4.1}) above implies that in this
case the operator $VU^{*}$ commutes with all density operators, since they
are unitary, it follows that $V=e^{i\theta }U$ for some real number $\theta $.

\section{A necessary condition}

In the notations and assumptions of the previous section, let us suppose
that the normalized state vector $\xi $ has the form with some constants $\left\{ t_{\mu }\right\} $ 
\begin{equation}
\xi =\sum_{\mu }t_{\mu }\varepsilon _{\mu }^{\prime \prime }\otimes
\varepsilon _{\mu }^{\prime }  \label{5.1}
\end{equation}
and let us look for the conditions under which the map (\ref{3.4}) becomes
linear. \bigskip 

\bigskip

\begin{proposition}
Given $\xi $ as in (\ref{5.1}) and $F$ as in (\ref{3.2}) the trace 
\[
tr_{123}(F(\rho \otimes |\psi \rangle \langle \psi |F)
\]
is independent of $\rho $ if and only if the coefficients $t_{k}$ of $\xi $
have the following form 
\begin{equation}
t_{k}=e^{i\theta _{k}}/\sqrt{N}  \label{5.2}
\end{equation}
In particular, if the condition (\ref{5.2}) above is satisfied, then the map
(\ref{3.4}) linearizes. 
\end{proposition}

\bigskip

\noindent 
{\bf Proof.} One has: 
\[
F(\rho \otimes |\psi \rangle \langle \psi |)F=\lambda _{h\alpha }\overline{\lambda }_{k\beta }F(\rho \otimes |\varepsilon _{h}^{\prime }\rangle \langle
\varepsilon _{k}^{\prime }|)F\otimes |\varepsilon _{\alpha }\rangle \langle
\varepsilon _{\beta }| 
\]
\[
=\lambda _{h\alpha }\overline{\lambda }_{k\beta }t_{\mu }\overline{t}_{\mu
^{\prime }}t_{\nu }\overline{t}_{\nu ^{\prime }}|\varepsilon _{\mu }^{\prime
\prime }\rangle |\varepsilon _{\mu }^{\prime }\rangle \langle \varepsilon
_{\mu ^{\prime }}^{\prime }|(\rho \otimes |\varepsilon _{h}^{\prime }\rangle
\langle \varepsilon _{k}^{\prime }|)|\varepsilon _{\nu }^{\prime \prime
}\rangle |\varepsilon _{\nu }^{\prime }\rangle \langle \varepsilon _{\nu
^{\prime }}^{\prime \prime }|\langle \varepsilon _{\nu ^{\prime }}^{\prime
}|\otimes |\varepsilon _{\alpha }\rangle \langle \varepsilon _{\beta }| 
\]
\[
=\lambda _{h\alpha }\overline{\lambda }_{k\beta }t_{\mu }\overline{t}_{\mu
^{\prime }}t_{\nu }\overline{t}_{\nu ^{\prime }}|\varepsilon _{\mu }^{\prime
\prime }\rangle \langle \varepsilon _{\mu ^{\prime }}^{\prime \prime },\rho
,\varepsilon _{\nu }^{\prime \prime }\rangle \langle \varepsilon _{\nu
^{\prime }}^{\prime \prime }|\otimes |\varepsilon _{\mu }^{\prime }\langle
\varepsilon _{\mu ^{\prime }}^{\prime },\varepsilon _{h}^{\prime }\rangle
\langle \varepsilon _{k}^{\prime },\varepsilon _{\nu }^{\prime }\rangle
\langle \varepsilon _{\nu ^{\prime }}^{\prime }|\otimes |\varepsilon
_{\alpha }\rangle \langle \varepsilon _{\beta }| 
\]
\[
=\lambda _{h\alpha }\overline{\alpha }_{k\beta }t_{\mu }\overline{t}_{h}t_{k} \overline{t}_{\nu },\langle \varepsilon _{h}^{\prime },\rho
\varepsilon _{k}^{\prime \prime }\rangle |\varepsilon _{\mu }^{\prime \prime
}\rangle \langle \varepsilon _{\nu ^{\prime }}^{\prime \prime }|\otimes
|\varepsilon _{\mu }^{\prime }\rangle \langle \varepsilon _{\nu ^{\prime
}}^{\prime }|\otimes |\varepsilon _{\alpha }\rangle \langle \varepsilon
_{\beta }| 
\]

Tracing over ${\cal H}_{1}\otimes {\cal H}_{2},$ 
\[
\lambda _{h\alpha }\overline{\lambda }_{k\beta }t_{\mu }\overline{t}_{h}t_{k} \overline{t}_{\mu }\langle \varepsilon _{h}^{\prime \prime },\rho
\varepsilon _{k}^{\prime \prime }\rangle |\varepsilon _{\alpha }\rangle
\langle \varepsilon _{\beta }| 
\]
Since $\sum_{\mu }|t_{\mu }|^{2}=1$, this is equal to 
\[
\lambda _{h\alpha }\overline{\lambda }_{k\beta }\overline{t}_{h}t_{k}\langle
\varepsilon _{h}^{\prime \prime },\rho \varepsilon _{k}^{\prime \prime
}\rangle |\varepsilon _{\alpha }\rangle \langle \varepsilon _{\beta }| 
\]

Taking the ${\cal H}_{3}$--trace, we find 
\[
\lambda _{h\alpha }\overline{\lambda }_{k\alpha }\overline{t}_{h}t_{k}\langle \varepsilon _{h}^{\prime \prime },\rho \varepsilon
_{k}^{\prime \prime }\rangle =|t_{h}|^{2}\langle \varepsilon _{h}^{\prime
\prime },\rho \varepsilon _{h}^{\prime \prime }\rangle 
\]
because of the unitarity of $\lambda _{\alpha ,\beta }$. \bigskip If follows
that the problem linearizes if and only if 
\[
\sum_{k}|t_{k}|^{2}|\varepsilon _{k}^{\prime \prime }\rangle \langle
\varepsilon _{k}^{\prime \prime }|=c=\mbox{constant}
\]
and this is equivalent to: 
\[
|t_{k}|^{2}=c\ ;\quad \forall \,k=1,\dots ,N
\]
Consequently $c=N$ and (\ref{5.2}) follows. \hfill 

\section{{Solution of the general teleportation problem: the case $N=2^{2m}$}
}

The remark at the end of section 2 and the result of section 4 show that, in
order to apply the solution of the weak teleportation problem to the
solution of the general one, one has to solve the the following problem:
find an o.n. basis $f_{n}$ of a finite dimensional Hilbert space ${\cal H}
$, identified to $\mathbf{C}^{N}$, such that one can determine the phases $
\sigma _{\alpha ,j}$ ($j=1,\ldots ,N$) so that the vectors 
\[
g_{\alpha }:={\frac{1}{N^{1/2}}}\sum_{j=1}^{N}e^{i\sigma _{\alpha ,j}}f_{j} 
\]
are still an o.n. basis of $\mathbf{C}^{N}$. If we further restrict the
condition requiring that $e^{i\sigma _{\alpha ,j}}=\pm 1$, then by
considering the two vectors $\psi _{\alpha }=\sum_{j=1}^{N}\varepsilon
_{j}^{\alpha }f_{j}$, $\psi _{\beta }=\sum_{j=1}^{N}\varepsilon _{j}^{\beta
}f_{j}$ with $\varepsilon _{j}^{\alpha },$ $\psi _{\beta }=\pm 1$, the
conditions 
\[
0=\langle \psi _{\alpha },\psi _{\beta }\rangle =\sum_{j}\varepsilon
_{j}^{\alpha }\varepsilon _{j}^{\beta } 
\]
show that a necessary condition for an affirmative answer to the above
problem is that $N$ is an even number. \bigskip

\bigskip

\begin{proposition}
\noindent  If $N=2^{m}$ for some $m\in \mathbf{N,}$ then the answer to the
above problem is affirmative.
\end{proposition}

\bigskip

\noindent 
{\bf Proof.} The statement is true for $m=1$. Assume by
induction that it is true for $m$ and consider $\mathbf{C}^{2^{2m+1}}\equiv 
\mathbf{C}^{2^{2m}}\otimes \mathbf{\ C}^{2}\equiv \mathbf{C}^{2^{m}}\oplus 
\mathbf{C}^{2^{m}}$.

Let $(\psi _{\alpha })$ $(\alpha =1,\dots ,2^{m})$ be an o.n. basis which
solves the problem for $\mathbf{C}^{2^{m}}$. Then clearly the set of vectors 
\[
{\frac{1}{\sqrt{2}}}\,\left( 
\begin{array}{l}
{\psi _{\alpha }} \\ 
{+\psi _{\alpha }}
\end{array}
\right) ,\quad {\frac{1}{\sqrt{2}}}\,\,\left( 
\begin{array}{l}
{\psi _{\alpha ^{\prime }}} \\ 
{-\psi _{\alpha ^{\prime }}}
\end{array}
\right) 
\]
is an o.n. basis of $\mathbf{C}^{2^{m+1}}$ in $Q_{2^{m+1}}$.

A more explicit solution of the problem is obtained as follows. We fix 
\[
N=2^{m} 
\]
and we choose 
\[
f_{\nu }:=e_{\nu _{1}}\otimes \dots \otimes e_{\nu _{m}}\ ;\quad \nu =(\nu
_{1},\dots ,\nu _{m})\in \{0,1\}^{m} 
\]
\[
\nu _{j}\in \{0,1\}\ ;\quad j=1,\dots ,m\ ;\quad e_{1}=\left( 
\begin{array}{l}
{1} \\ 
{0}
\end{array}
\right) ,\ e_{0}=\left( 
\begin{array}{l}
{0} \\ 
{1}
\end{array}
\right) 
\]

We know that 
\begin{equation}
g_{0}:={\frac{1}{\sqrt{2}}}\,(e_{1}+e_{0})={\frac{1}{\sqrt{2}}}\,\left( 
\begin{array}{l}
{1} \\ 
{1}
\end{array}
\right) ;\quad g_{1}:={\frac{1}{\sqrt{2}}}\,(e_{1}-e_{0})={\frac{1}{\sqrt{2}}}\,\left( 
\begin{array}{l}
{1} \\ 
-{1}
\end{array}
\right) 
\end{equation}
is an o.n. basis of $\mathbf{C^{2}}$ so that 
\begin{equation}
g_{\alpha }:=g_{\alpha _{1}}\otimes \dots \otimes g_{\alpha _{m}}\quad
;\qquad \alpha :=(\alpha _{1},\dots ,\alpha _{m})\in \{0,1\}^{m}
\end{equation}
is an o.n. basis of $\bigotimes\limits_{1}^{m}\mathbf{C}^{2}=\mathbf{C}^{2^{m}}$. However we have 
\begin{eqnarray}
g_{\alpha _{1}}\otimes \dots \otimes g_{\alpha _{m}} &=&{\frac{1}{2^{m/2}}}\,\bigotimes\limits_{j=1}^{m}(e_{1}+(-1)^{\alpha _{j}}e_{0})  \nonumber \\
{} &=&{\frac{1}{2^{m/2}}}\,\sum_{\nu _{1},\dots ,\nu
_{m}}(-1)^{\sum\limits_{j=1}^{m}(1-\nu _{j})\alpha _{j}}e_{\nu _{1}}\otimes
\dots \otimes e_{\nu _{m}}  \nonumber \\
{} &=&{\frac{1}{2^{m/2}}}\,\sum_{\nu \in \{0,1\}^{m}}(-1)^{\sigma _{\alpha
}(\nu )}e_{\nu }=g_{\alpha }\quad \quad \quad \quad \quad \quad \quad 
\label{6.3}
\end{eqnarray}
where 
\[
\sigma _{\alpha }(\nu ):=\sum_{j=1}^{m}(1-\nu _{j})\alpha _{j}.\hfill 
\]

In the notations of section 3, let $N=2^{m}$ for some $m\in \mathbf{N}$ and
let the orthonormal bases $(\varepsilon _{\alpha }^{\prime })$ of ${\cal H}_{2}$ and $(\varepsilon _{\alpha }^{\prime \prime })$ of ${\cal H}_{1}$
in (\ref{3.6}), (\ref{3.7}) be two copies of the basis $e_{\nu _{1}}\otimes
\dots \otimes e_{\nu _{m}}$, described in Proposition 6.1. Let $(\varepsilon
_{\alpha })$ be an arbitrary orthonormal basis of ${\cal H}_{3}$ and let $\psi $ and $U$ be as in Proposition 3.1. Then, it is easy to see the
followimg corollary.

\bigskip

\begin{corollary}
\noindent  If the vector $\xi _{\alpha }$ for each $\alpha \in \{0,1\}^{m}$
is defined by 
\begin{equation}
\xi _{\alpha }:={\frac{1}{2^{m/2}}}\,\sum_{\mu \in \{0,1\}^{m}}(-1)^{\sigma
_{\alpha }(\nu )}\varepsilon _{\mu }^{\prime }\varepsilon _{\mu }^{\prime
\prime }
\end{equation}
then, for each $\alpha \in \{0,1\}^{m}$, the triple $\{\psi ,U,\xi _{\alpha
}\}$ solves the weak teleportation problem and the projections 
\[
F_{\alpha }:=|\xi _{\alpha }\rangle \langle \xi _{\alpha }|\otimes 1_{3}
\]
are mutually orthogonal. 
\end{corollary}

\medskip

\begin{remark}
\noindent  In the above construction, nothing prevents the possibility of
choosing a different unitary matrix $(\lambda _{j,k})$ for each $\alpha \in
\{0,1\}^{m}$ so that the triple $\{\psi _{\alpha },U_{\alpha },\xi _{\alpha
}\}$ solves the teleportation problem in the general formulation of section
2. 
\end{remark}

We here notice that the BBCJPW scheme provides nice examples to our results.
In their scheme, $\sigma ^{(23)}$ is given by an EPR spin pair in a singlet
state such as 
\[
\sigma ^{(23)}=|\psi \rangle \langle \psi |, 
\]
where

\[
|\psi \rangle =c|\uparrow ^{(2)}\rangle \otimes |\downarrow ^{(3)}\rangle
+d|\downarrow ^{(2)}\rangle \otimes |\uparrow ^{(3)}\rangle ,\quad \left|
c\right| ^{2}+\left| d\right| ^{2}=1. 
\]
with the spin up vector $|\uparrow \rangle :=$ $e_{1}=\left( 
\begin{array}{l}
{1} \\ 
{0}
\end{array}
\right) $ and the spin down vector $|\downarrow \rangle :=e_{0}=\left( 
\begin{array}{l}
{0} \\ 
{1}
\end{array}
\right) .$ There, Alice's measurement $F_{k}^{(12)}$ is chosen in the
partition of the identity $\left\{ F_{k}^{(12)};k=1,2,3,4\right\} $ ;

\begin{eqnarray*}
F_{1}^{(12)} &=&|\xi ^{(-)}\rangle \langle \xi ^{(-)}|,\ F_{2}^{(12)}=|\xi
^{(+)}\rangle \langle \xi ^{(+)}|,\  \\
F_{3}^{(12)} &=&|\zeta ^{(-)}\rangle \langle \zeta ^{(-)}|,\
F_{4}^{(12)}=|\zeta ^{(+)}\rangle \langle \zeta ^{(+)}|
\end{eqnarray*}
with 
\begin{eqnarray}
|\xi ^{(-)}\rangle &=&\sqrt{\frac{1}{2}}\left( |\uparrow ^{(1)}\rangle
\otimes |\downarrow ^{(2)}\rangle -|\downarrow ^{(1)}\rangle \otimes
|\uparrow ^{(2)}\rangle \right)  \nonumber \\
|\xi ^{(+)}\rangle &=&\sqrt{\frac{1}{2}}\left( |\uparrow ^{(1)}\rangle
\otimes |\downarrow ^{(2)}\rangle +|\downarrow ^{(1)}\rangle \otimes
|\uparrow ^{(2)}\rangle \right)  \nonumber \\
|\zeta ^{(-)}\rangle &=&\sqrt{\frac{1}{2}}\left( |\uparrow ^{(1)}\rangle
\otimes |\uparrow ^{(2)}\rangle -|\downarrow ^{(1)}\rangle \otimes
|\downarrow ^{(2)}\rangle \right)  \nonumber \\
|\zeta ^{(+)}\rangle &=&\sqrt{\frac{1}{2}}\left( |\uparrow ^{(1)}\rangle
\otimes |\uparrow ^{(2)}\rangle +|\downarrow ^{(1)}\rangle \otimes
|\downarrow ^{(2)}\rangle \right) .  \nonumber
\end{eqnarray}

The unitary (key) operators $U_{k}$ ($k=1,2,3,4)$ are given as follows 
\[
\begin{array}{l}
U_{1}\equiv |\uparrow ^{(1)}\rangle \langle \uparrow ^{(3)}| +|\downarrow
^{(1)}\rangle \langle \downarrow ^{(3)}| \\ 
U_{2}\equiv |\uparrow ^{(1)}\rangle \langle \uparrow ^{(3)}| -|\downarrow
^{(1)}\rangle \langle \downarrow ^{(3)}| \\ 
U_{3}\equiv |\uparrow ^{(1)}\rangle \langle \downarrow ^{(3)}| +|\downarrow
^{(1)}\rangle \langle \uparrow ^{(3)}| \\ 
U_{4}\equiv |\uparrow ^{(1)}\rangle \langle \downarrow ^{(3)}| -|\downarrow
^{(1)}\rangle \langle \uparrow ^{(3)}|.
\end{array}
\]

In \cite{BBCJPW}, they discussed a more general example in which $\sigma_{23}=|\psi\rangle \langle \psi|$ $\in {\cal S}({\cal H}_{2}\otimes {\cal H}_{3})$ with ${\cal H}_{2}={\cal H}_{3})= 
\mathbf{C}^N$ and 
\[
|\psi \rangle =\sum_{j=0}^{N-1}\frac{1}{\sqrt{N}}|j\rangle \otimes |j\rangle
.\quad 
\]
In this case, Alice's measurement is performed with projections of the form

\[
F_{nm}^{(12)}=|\xi _{nm}\rangle \langle \xi _{nm}| 
\]
where 
\[
|\xi _{nm}\rangle =\frac{1}{\sqrt{N}}\sum_{j=0}^{N-1}e^{\frac{2\pi ijn}{N}}|j\rangle \otimes |\left( j+m\right) \ \mathrm{mod}\ N\rangle . 
\]

These states are examples of the entangled states and Alice's projections in
our general teleportation scheme.

Finally we note that the channels constructed by the above states are
linear, and they could show the teleportation of the initial state attached
to ${\cal H}_{1}$.

\end{document}